\documentclass{cernart}
\usepackage{epsfig}

\newcommand{\beq}  {\begin{equation}}
\newcommand{\eeq}  {\end{equation}}
\newcommand{\bmath}{\begin{eqnarray}}
\newcommand{\emath}{\end{eqnarray}}
\newcommand{\bei}{\begin{itemize}}
\newcommand{\eei}{\end{itemize}}
\newcommand{\ks}{K_S}
\newcommand{\kl}{K_L}
\newcommand{\gaga}{\gamma\gamma}

\newcommand{\pgg}{\pi^0\gamma\gamma}
\newcommand{\kspgg}{K_{S}\rightarrow\pi^0\gamma\gamma}
\newcommand{\kspee}{K_{S}\rightarrow\pi^0e^+e^-}
\newcommand{\klpgg}{K_{L}\rightarrow\pi^0\gamma\gamma}

\newcommand{\klgg}{K_{L}\rightarrow\gamma\gamma}
\newcommand{\kspp}{K_{S}\rightarrow{\pi^0}{\pi^0}}
\newcommand{\ksppd}{K_{S}\rightarrow{\pi^0}{\pi^0_D}}
\newcommand{\klppp}{K_{L}\rightarrow{\pi^0}{\pi^0}{\pi^0}}

\newcommand{\pipin}{{\pi^0}{\pi^0}}

\newcommand{\eoe}{Re({\epsi'}/{\epsi})}
\newcommand{\epsi}{\varepsilon}
\newcommand{\chpt}{${\chi}PT$}

\begin{document}
\begin{titlepage}
\docnum{CERN--EP/2003-052}
\date{28 August 2003}

\vspace{3cm}

\title{\bf \LARGE First observation of the $\kspgg$ decay}
\collaboration{\Large NA48 Collaboration}

\vspace{2cm}

\begin{abstract}
Using the NA48 detector at the CERN SPS, 31 $\kspgg$ candidates with
an estimated background of $13.7 \pm 3.2$ events
have been observed. This first observation leads to a
branching ratio of $BR(\kspgg) = (4.9 \pm 1.6_{stat} \pm 0.9_{syst})
\times 10^{-8}$ in agreement with Chiral Perturbation theory predictions.
\end{abstract}

\vfill

\submitted{Submitted to Phys. Lett. B}

\newpage

\begin{Authlist}
\begin{center}
{\bf NA48 collaboration}
\  \\[0.2cm] 
 A.~Lai,
 D.~Marras

Dipartimento di Fisica dell'Universit\`a e Sezione dell'INFN di Cagliari, I-09100 Cagliari, Italy \\[0.2cm] 
J.R.~Batley,
 R.S.~Dosanjh,
 T.J.~Gershon,
G.E.~Kalmus,
 C.~Lazzeroni,
 D.J.~Munday,
E.~Olaiya,
 M.A.~Parker,
 T.O.~White,
 S.A.~Wotton

Cavendish Laboratory, University of Cambridge, Cambridge, CB3 0HE, U.K.\footnote{ Funded by the U.K.    Particle Physics and Astronomy Research Council}
 \\[0.2cm] 
R.~Arcidiacono\footnote{ On leave from Dipartimento di Fisica Sperimentale 
dell'Universit\`a e    Sezione dell'INFN di Torino,  I-10125 Torino, Italy}
,
G.~Barr\footnote{Present address: University of Oxford, Keble Road,
  Oxford, OX1 3RH, U.K.}
,
 G.~Bocquet,
 A.~Ceccucci,
 T.~Cuhadar-D\"onszelmann,
 D.~Cundy\footnote{Present address: Istituto di Cosmogeofisica del CNR
  di Torino, I-10133 Torino, Italy}
,
 N.~Doble\footnote{Also at Dipartimento di Fisica dell'Universit\`a e
  Sezione dell'INFN di Pisa, I-56100 Pisa, Italy}
,
V.~Falaleev,
 L.~Gatignon,
 A.~Gonidec,
 B.~Gorini,
 P.~Grafstr\"om,
W.~Kubischta,
 A.~Lacourt,
A.~Norton,
 B.~Panzer-Steindel,
G.~Tatishvili\footnote{ On leave from Joint Institute for Nuclear Research, Dubna,141980, Russian Federation}
,
 H.~Wahl\footnote{Also at Dipartimento di Fisica dell'Universit\`a e
  Sezione dell'INFN di Ferrara, I-441000 Ferrara, Italy}

CERN, CH-1211 Gen\`eve 23, Switzerland \\[0.2cm] 
C.~Cheshkov,
 P.~Hristov,
V.~Kekelidze,
 D.~Madigojine,
N.~Molokanova,
Yu.~Potrebenikov,
 A.~Zinchenko,
Joint Institute for Nuclear Research, Dubna, Russian    Federation \\[0.2cm] 
%
%
 V.~Martin,
 P.~Rubin\footnote{On leave from the University of Richmond, Richmond,
 VA, 23173, USA; Funded in part by the US National Science Foundation
 under grant 9971970}
,
 R.~Sacco,
 A.~Walker

Department of Physics and Astronomy, University of    Edinburgh, JCMB King's Buildings, Mayfield Road, Edinburgh,    EH9 3JZ, U.K. \\[0.2cm] 
M.~Contalbrigo,
 P.~Dalpiaz,
 J.~Duclos,
P.L.~Frabetti,
 A.~Gianoli,
 M.~Martini,
 F.~Petrucci,
 M.~Savri\'e

Dipartimento di Fisica dell'Universit\`a e Sezione    dell'INFN di Ferrara, I-44100 Ferrara, Italy \\[0.2cm] 
%
%
A.~Bizzeti\footnote{ Dipartimento di Fisica
dell'Universita' di Modena e Reggio Emilia, via G. Campi 213/A
I-41100, Modena, Italy}
,
M.~Calvetti,
 G.~Collazuol,
 G.~Graziani,
 E.~Iacopini,
 M.~Lenti,
 F.~Martelli\footnote{Istituto di Fisica, Universit\'a di Urbino, I-61029
  Urbino, Italy}
,
 M.~Veltri\footnotemark[10]

Dipartimento di Fisica dell'Universit\`a e Sezione    dell'INFN di Firenze, I-50125 Firenze, Italy \\[0.2cm] 
%
%
 M.~Eppard,
 A.~Hirstius,
 K.~Holtz,
 K.~Kleinknecht,
 U.~Koch,
 L.~K\"opke,
 P.~Lopes da Silva, 
P.~Marouelli,
 I.~Mestvirishvili,
 I.~Pellmann,
 A.~Peters,
S.A.~Schmidt,
  V.~Sch\"onharting,
 Y.~Schu\'e,
 R.~Wanke,
 A.~Winhart,
 M.~Wittgen

Institut f\"ur Physik, Universit\"at Mainz, D-55099    Mainz, Germany\footnote{ Funded by the German Federal Minister for    Research and Technology (BMBF) under contract 7MZ18P(4)-TP2}
 \\[0.2cm] 
J.C.~Chollet,
 L.~Fayard,
 L.~Iconomidou-Fayard,
 G.~Unal,
 I.~Wingerter-Seez

Laboratoire de l'Acc\'el\'erateur Lin\'eaire,  IN2P3-CNRS,Universit\'e de Paris-Sud, 91898 Orsay, France\footnote{ Funded by Institut National de Physique des
 Particules et de Physique Nucl\'eaire (IN2P3), France}
 \\[0.2cm] 
G.~Anzivino,
 P.~Cenci,
 E.~Imbergamo,
 P.~Lubrano,
 A.~Mestvirishvili,
 A.~Nappi,
M.~Pepe,
 M.~Piccini

Dipartimento di Fisica dell'Universit\`a e Sezione    dell'INFN di Perugia, I-06100 Perugia, Italy \\[0.2cm] 
%
%
R.~Casali,
 C.~Cerri,
 M.~Cirilli\footnote{Present address: Dipartimento di Fisica
 dell'Universit\'a di Roma ``La Sapienza'' e Sezione dell'INFN di Roma,
 00185 Roma, Italy}
,
F.~Costantini,
 R.~Fantechi,
 L.~Fiorini,
 S.~Giudici,
 I.~Mannelli,
G.~Pierazzini,
 M.~Sozzi

Dipartimento di Fisica, Scuola Normale Superiore e Sezione dell'INFN di Pisa, I-56100 Pisa, Italy \\[0.2cm] 
%
%
J.B.~Cheze,
 M.~De Beer,
 P.~Debu,
 F.~Derue,
 A.~Formica,
 R.~Granier de Cassagnac,
G.~Gouge,
G.~Marel,
E.~Mazzucato,
 B.~Peyaud,
 R.~Turlay,
 B.~Vallage

DSM/DAPNIA - CEA Saclay, F-91191 Gif-sur-Yvette, France \\[0.2cm] 
M.~Holder,
 A.~Maier,
 M.~Ziolkowski 

Fachbereich Physik, Universit\"at Siegen, D-57068 Siegen, Germany\footnote{ Funded by the German Federal Minister for Research and Technology (BMBF) under contract 056SI74}
 \\[0.2cm] 
 C.~Biino,
 N.~Cartiglia,
 F.~Marchetto, 
E.~Menichetti,
 N.~Pastrone

Dipartimento di Fisica Sperimentale dell'Universit\`a e    Sezione dell'INFN di Torino,  I-10125 Torino, Italy \\[0.2cm] 
J.~Nassalski,
 E.~Rondio,
 M.~Szleper,
 W.~Wislicki,
 S.~Wronka

Soltan Institute for Nuclear Studies, Laboratory for High    Energy Physics,  PL-00-681 Warsaw, Poland\footnote{Supported by the Committee for Scientific Research grants
5P03B10120, SPUB-M/CERN/P03/DZ210/2000 and SPB/CERN/P03/DZ146/2002}
 \\[0.2cm] 
H.~Dibon,
 M.~Jeitler,
 M.~Markytan,
 I.~Mikulec,
 G.~Neuhofer,
M.~Pernicka,
 A.~Taurok,
 L.~Widhalm

\"Osterreichische Akademie der Wissenschaften, Institut  f\"ur Hochenergiephysik,  A-1050 Wien, Austria\footnote{    Funded by the Austrian Ministry for Traffic and Research under the    contract GZ 616.360/2-IV GZ 616.363/2-VIII, and by the Fonds f\"ur    Wissenschaft und Forschung FWF Nr.~P08929-PHY}
 \\[0.2cm] 
\end{center}

\end{Authlist}
\setcounter{footnote}{0}

\end{titlepage}

\section{\bf Introduction}

Radiative non-leptonic rare kaon decays have proved to be useful tests
for low energy hadron dynamics studied in the framework of 
the Standard Model
by the Chiral Perturbation theory (\chpt). This paper describes 
the first observation of the $\kspgg$ decay obtained from data taken by 
the NA48 experiment in the year 2000.

In the decay $\kspgg$, the photon pair production is described by an
amplitude dominated by a pseudo-scalar
meson pole. In \chpt\ this pole is dominated by $\pi^0$
contribution, and the lowest order amplitude is non-vanishing, in contrast to
the similar $\klgg$ decay. 
The theoretical prediction for the branching ratio 
is $3.8 \times 10^{-8}$ with higher order corrections expected to be
small~\cite{epd}
and is quoted in the kinematic region
$z=m_{\gaga}^2/m_K^2~>~0.2$ which is free from the overwhelming $\kspp$
background. A measurement of the branching ratio can provide
information about the presence of 
extra non-pole contributions studied e.g. in~\cite{bpp}.
In addition, the momentum dependence
of the weak vertex which is predicted by
\chpt\ can be tested by the measured shape of the $z$ spectrum.
The lowest previously published limit on the branching ratio is
$BR(\kspgg)_{z>0.2} < 3.3 \times 10^{-7}$ at 90\% confidence 
level~\cite{lucia}.

\section{\bf Experimental set-up and data taking}
 
The NA48 detector was designed to measure direct CP violation
in the decays of $\kl$ and $\ks$ into
$\pi\pi$, described by the parameter
$\eoe$, using simultaneous far- and near-target beams~\cite{doble}. 
The analysis presented in this paper
is based on data recorded 
during a special 40-day run performed in 
the year 2000 with near-target beam only and with an
intensity of about $\sim 10^{10}$ 400 GeV protons hitting the target 
during the 
3.2 s long SPS spill. The exit of the collimator, 6 m downstream of the
target, is followed by a wide vacuum tube approximately 113 m long
containing the decay region and
terminated by an aluminium window close to the
detector\footnote{In the usual NA48 experimental configuration the
  vacuum tube is terminated at 89 m by a thin Kevlar window 
 followed by a helium filled tank which
  contains drift chambers and a spectrometer magnet. For these data
  however, the Kevlar window and the
  drift chambers were removed, the helium tank was evacuated and the
  magnet was switched off.}.

The detector elements used in the analysis are the following:
\bei
\item A liquid krypton calorimeter (LKr)~\cite{unal}, placed less than
  2 m behind the aluminium window, is used to
  measure the energy, position and time of
electro-magnetic showers. 
The energy resolution is
\beq
\sigma(E)/E \simeq 0.090/E \oplus 0.032/\sqrt{E} \oplus 0.0042, \nonumber
\eeq
where E is in GeV. 
The position and time resolutions for a single photon with energy
larger than 20 GeV are better than 1.3 mm
and 300 ps, respectively.
\item A sampling hadron calorimeter composed of 48 steel plates, 
interleaved with
scintillator planes, with a readout in horizontal and
vertical projections.
\item Seven ring shaped scintillator counters equipped with
  iron converters (AKL), used to detect photons escaping the
  outer limits of the calorimeter acceptance.
\eei
The trigger decision,
common to $\pi^0\gamma\gamma$ and $\pipin$ decays, is
based on quantities which are derived from the projections 
of the energy deposited in the
electro-magnetic liquid krypton calorimeter~\cite{nutnim}. 
The trigger required that the total deposited 
energy $E_{tot}$ is larger than 50 GeV, the radial distance of the centre of
energy from the beam axis is smaller than 15 cm and the proper life time of the kaon
is less than 9 $\ks$ lifetimes downstream of the collimator. The
inefficiency of the trigger was measured to be at the level of 0.1\%.

More about the detector and
the experimental configuration
during data taking in the year 2000 can be found in~\cite{ksgg}. 

\section{\bf Event selection }

The energies and positions of electro-magnetic showers initiated by
photons, measured in 
the liquid krypton calorimeter, are used to 
calculate the kaon energy and decay vertex. To select $\kspgg$ and
$\kspp$ candidates,
all events with $\geq 4$ energy clusters are considered. All
combinations of four
clusters which are within 5 ns from the average time and with no other 
cluster with energy $> 1.5$ GeV closer in time than 7 ns with respect
to the event time  are selected. The event time is computed from the
times of the most energetic cells of the selected clusters.
In addition, the clusters must pass the following cuts:
\begin{itemize}
   \item The energy of each cluster must be greater than 3 GeV
         and less than 100 GeV.
   \item The transverse distance between two clusters is required to 
         be greater than 10 cm.
   \item The total energy of the selected cluster combination 
         is required to be 
         less than 170 GeV and greater than 70 GeV.
   \item The distance of the centre of energy from the beam axis, 
\beq       
R_C = \frac{\sqrt{(\sum_i E_i x_i)^2+(\sum_i E_i y_i)^2}}
            {\sum_i E_i}
\eeq
         is required to be less than 5 cm, where $E_{i}$,
$x_{i}$, $y_{i}$ are the $i$-th cluster energy, 
$x$ and $y$ coordinates at LKr, respectively.
   \item The energy deposited in the hadron calorimeter must not
         exceed 3 GeV in a time window of $\pm15$ ns around the event time.
   \item The AKL counter should not register any hit in a time window of $\pm7$ ns around
   the event time
\end{itemize} 

The decay vertex position $z_{vertex}$ of a kaon 
is calculated using the kaon mass constraint, $m_K$,
\beq       
z_{vertex}= z_{Lkr} - \frac{\sqrt{\sum_{i,j>i}{E_i}{E_j}
                   d^2_{ij}}}{m_K}
\eeq

where $z_{Lkr}$ is the longitudinal 
coordinate of the LKr calorimeter with respect
to the end of the collimator and $d_{ij}$ is the transverse distance between
$i$-th and $j$-th cluster at the calorimeter. The invariant mass of
photon pairs is calculated using $z_{vertex}$.

The $\kspgg$ candidates must have a photon pair with an invariant mass 
within 2~MeV/c$^2$ of the $\pi^0$ mass, $m_{\pi^0}$, and
the other pair must satisfy the condition $m_{\gaga}^2/m_K^2 > 0.2$.
In the $\kspp$ sample a
$\chi^2$ cut of 27 ($\sim{5\sigma}$) is applied to the invariant
masses of photon pairs, 
defined as
\beq
 \chi^2=\left[ \frac{(m_1+m_2)/2-m_{\pi^0}}{\sigma_+}\right]^2+
        \left[ \frac{(m_1-m_2)/2}{\sigma_-}\right]^2
\eeq
where $\sigma_{\pm}$ are the resolutions on $(m_1 \pm m_2)/2$ measured
from the data and parametrised as a function of the lowest photon
energy. The typical value of $\sigma_+$ is 0.4~MeV/c$^2$ and 
$\sigma_-$ is 0.8~MeV/c$^2$~\cite{eps}.
For $\kspgg$ candidates the $\chi^2$ is required to be larger than 5400.

\begin{figure}[ht]
\begin{center}
\mbox{\epsfig{file=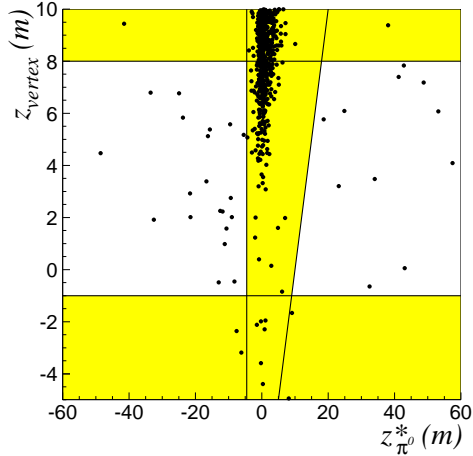,width=6.5cm}}
\end{center}
\caption{Illustration of $z_{vertex}$ and $z^*_{\pi^0}$ cuts on the
  data. All
  other cuts were applied. Rejected
areas are shaded. Both variables are defined with respect to the
exit of the collimator. The accumulation of events around small
$z^*_{\pi^0}$ values is due to $\ksppd$ background at high
$z_{vertex}$ and due mainly to pile-up events at low or negative $z_{vertex}$.}
\label{fig:zstplot}
\end{figure}

In order to suppress the large
background from $\ksppd$ decays with one particle escaping the
acceptance the decay region was restricted
to be between -1 m and 8 m with respect to the exit of the
collimator. Most of the $\ksppd$ background events have an apparent
$z_{vertex}$ downstream of this region because of the missing energy. This
condition is even more effective against background from $\klppp$.
The $\ksppd$ background is further suppressed by imposing the $\pi^0$
mass hypothesis on the remaining four
pairings\footnote{If indexes 1 and 2 refer to $\gamma$'s assigned to
  the $\pi^0$ then these four pairings are 1-3, 1-4, 2-3 and 2-4.}
of photon showers and calculating the vertex
positions $z_{\pi^0}$ for each of these $\gamma$ pairs.
Denoting as $z^*_{\pi^0}$ the closest $z_{\pi^0}$
to the exit of the collimator, events with
\beq
\label{zstar}
z^*_{\pi^0}<-4.5\mathrm{ m}
\quad
\mathrm{or}
\quad
z^*_{\pi^0}>z_{vertex}+10\mathrm{ m}
\eeq
are accepted (Fig.~\ref{fig:zstplot}). This cut is very effective 
in suppressing $\ksppd$ background because, as shown by Monte Carlo
studies, this background is completely dominated by
events with one of the $\pi^0_D \rightarrow ee\gamma$ decay products 
escaping detection. 
In addition, this cut removes events from inelastic scattering
of beam particles in the collimator with multiple $\pi^0$
production. The downstream end of this cut is
intentionally extended in order to reduce
background from pile-up of a $\pi^0$ from $\kspp$ with $\gamma$'s from
another decay.

A small amount of $\ksppd$
background survives the 
above cuts if a $\gamma$ from one
$\pi^0$ and a $\gamma$ or an $e$ from the other $\pi^0$ overlap. 
In this case the
reconstructed $z_{vertex}$ does not move downstream because the entire
kaon energy is collected in the LKr calorimeter 
and there is no pair of showers which would
give a correct $z_{\pi^0}$ under a $\pi^0$ hypothesis. Still, by
assuming this type of overlap the correct vertex position can be
reconstructed by
\beq
\label{overlap}
z_{overlap}=z_{LKr}-\frac{1}{m_{\pi^0}}
\sqrt{
E_1 d^2_{14} 
\frac{\sum_{j>1,i>j}^{4} E_i E_j d^2_{ij}}
{\sum_{i}^{3} E_i d^2_{i4}}
}
\eeq
where index 1 refers to the $\gamma$ from the $\pi^0$, indexes 2 and 3 to
$\gamma$ or e's
from $\pi^0_D$ and index 4 to the overlapping shower.
For $\pgg$ events $z_{overlap} \neq z_{vertex}$ and hence
a cut $|z_{overlap}-z_{vertex}|>1.5$ m on all 12 possible combinations 
reduces this type of background to a sufficiently low level, without
significant loss of signal.

In order to remove events with hadrons or 
overlapping showers from the $\pi^0\gamma\gamma$ sample, 
the shower width is required to be less than
3$\sigma$ above the average value for photon showers of a given energy. 
This cut, which
is calibrated using showers in $\kspp$ decays, 
removes $<1\%$ of good $\kspgg$ events. 

In addition, residual
background from $\Xi^0 \rightarrow \Lambda \pi^0$ with subsequent
$\Lambda \rightarrow n \pi^0$ decay, where one $\gamma$ escapes and the
neutron produces a photon-like shower
in the LKr calorimeter, is suppressed by exploiting the large momentum
asymmetries in both $\Xi^0$ and $\Lambda$ decays and accepting only
events with: 
\beq
\label{asym}
\frac{|E_3-E_4|}{E_3+E_4}<0.35
\quad
\mathrm{and}
\quad
\frac{(E_3+E_4)-(E_1+E_2)}{E_1+E_2+E_3+E_4}<0.3
\eeq
where indexes 1 and 2 refer to the shower pair identified as
photons from the 
$\pi^0$ and 3, 4 to other two showers.

The background from accidental pile-up events is reduced by strengthening the
requirements on shower times. Two configurations of a pile-up are
considered: 3+1 showers and 2+2 showers. The 3+1 configuration is best
described by a variable
\beq
\label{t3max}
t_{3max} = [ t_i - {\sum_{j\neq i} t_j}/{3} ]_{max}
\eeq
where the difference between any shower time and the average of the other
three is maximised. Similarly for the 2+2 configuration one can use a
variable
\beq
\label{t2max}
t_{2max} = [ t_i - {\sum_{j\neq i} t_j}/{2} ]_{max}
\eeq
in which only two of three remaining shower times are averaged. For the
3+1 pile-up topology $t_{2max}$ and $t_{3max}$ are always equal within the
time resolution, however
in the 2+2 case the two variables have different values in principle. This
allows one to analyse this type of background in more detail. 
For the selection of the
signal sample a single cut on $t_{2max}$ of $<$1~ns is chosen, 
because $t_{2max}$ describes both configurations
in the same way.

After all selection cuts 31 events remain in the sample. For the
normalisation a sample scaled-down by factor 1000 of
about $285\times 10^3$ $\kspp$ events has been used.

\section{\bf Background subtraction}

The following processes have been considered as potential sources of
residual background:
\begin{itemize}
\item irreducible $\klpgg$ background from the $\kl$ component of the beam 
\item $\kspp$ with mis-reconstructed energy 
\item $\ksppd$ with a $\gamma$ or an $e$ escaping detection
  or with a shower overlap
\item Hadron background, mainly
$\Xi^0 \rightarrow \Lambda \pi^0$ with subsequent
$\Lambda \rightarrow n \pi^0$, with three photon showers and one narrow
  neutron shower in the LKr calorimeter
\item Pile-up of two decays where the two decaying particles originate
  at the target 
  either from two protons (accidental pile-up) or from 
a single proton (in-time pile-up).
\end{itemize}

The amount of $\klpgg$ admixture in the signal sample is estimated
using the $\kl$ flux measured from the $\kspp$ rate assuming equal production
of $\ks$ and $\kl$ at the target. The acceptance was calculated 
using a Monte Carlo simulation. This background amounts to $3.8 \pm 0.2$ events.

$\kspp$ events can pass the cuts on invariant masses, and
especially $\chi^2 > 5400$, only if far
non-Gaussian tails are present in the energy reconstruction from the LKr
calorimeter. In order to study cases where
one of the four shower energies may be
mis-reconstructed, making use of the over-constraint of $\kspp$
events, the invariant mass of the event is reconstructed by using only
three out of four shower energies and all four positions. None of the
signal events has
been found to have an invariant mass close to the kaon mass in any of
the shower combinations. In addition, a toy Monte Carlo was employed 
to generate
$\kspp$ events using realistic resolutions and non-Gaussian tails in
energy, enhanced by an order of magnitude with respect to 
those known from E/p studies
of electrons from the $\eoe$ analysis. In a sample equivalent to 
twice the flux
of the 2000 near-target run 
no event passes the $\chi^2$,
$m_{\gaga}$ and $z_{vertex}$ cuts at the same time.

The residual $\ksppd$ background has been studied using
Monte Carlo simulations. Generating 1.7 times the collected flux, 4 events
have been found to pass all cuts. Three of them have
overlapping showers and reside in the tail of the $z_{overlap}-z_{vertex}$ 
distribution. The fourth event decays in the collimator, and one of the
photons undergoes a conversion in the collimator walls, while
the $e^+e^-$ pair from the $\pi^0$
passes trough the central hole of the detector. Scaling these events to the observed
kaon flux, a background estimate of $2.4\pm 1.2$ events is obtained with an
uncertainty dominated by the statistics. The systematic error of this
estimate has been checked by relaxing $z^*_{\pi^0}$ and $z_{overlap}$
cuts. The number of events and their $z_{vertex}$ distribution agree
with data within few percent.

The amount of hadron background surviving cuts on energy deposited in
the hadron
calorimeter, the shower width and the energy asymmetry (\ref{asym}) 
was estimated
by extrapolating the shower width distribution from large widths to
the signal region. The shower width distribution of neutrons was extracted
from fully reconstructed 
$\Xi^0 \rightarrow \Lambda \pi^0 \rightarrow n \pi^0\pi^0$
events. An estimate of $0.1\pm 0.1$ events has been obtained.

The accidental pile-up background has been studied using
time variables $t_{2max}$ (\ref{t2max}) and  $t_{3max}$ (\ref{t3max})
in a time window of 6 ns which is 
six times larger than the signal time window. In this time window the event
distribution is not affected by any selection cut. It turns out
that about 50\% of the background is in a 3+1 configuration while 50\% is
2+2. Since both of these configurations are described equivalently by
$t_{2max}$ this variable is used to extrapolate from the control
region $2<|t_{2max}|<6$ ns to the signal region $|t_{2max}|<1$ ns 
assuming a flat distribution. This extrapolation gives an estimate of $7.0
\pm 1.3$ events for the accidental pile-up background. The order of
magnitude of this estimate was confirmed by overlaying 3-photon events
from a $\kspp$ toy Monte Carlo with random showers detected close in time
(20-30 ns) to fully reconstructed $\kspp$ events 
and by taking into
account the kaon flux.

\begin{figure}[ht]
\begin{center}
\mbox{\epsfig{file=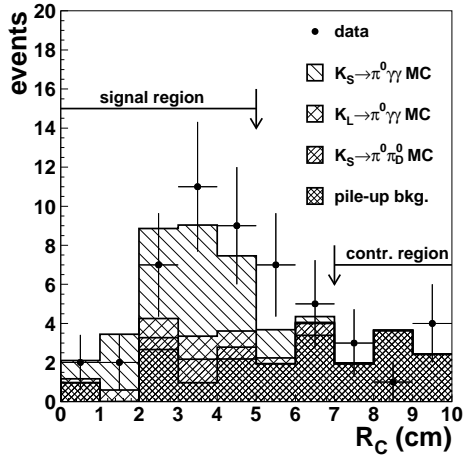,width=6.5cm}}
\end{center}
\caption{The $R_C$ distribution of data compared to the sum of background
  models and signal Monte Carlo scaled by the calculated branching
  ratio. The control region for pile-up background subtraction is indicated.}
\label{fig:rcplot}
\end{figure}

The in-time pile-up background in general has a  topology similar to the
accidental pile-up, but, since it is generated by the same proton in
the target, its rate cannot be measured by extrapolating from the out-of-time
sample. The relative rate of in-time and accidental pile-up has been
studied using fully reconstructed $\kspp$ with additional showers in
the LKr calorimeter. It has been found that a 20\% enhancement of
pile-up background due to an in-time component cannot be
excluded. In order to estimate the amount of total pile-up background
without using time variables, the distribution of
$R_C$ has been employed. The background is extrapolated from a control region
$7<R_C<10$ cm to the signal region $R_C<5$ cm by using the $R_C$ shape
obtained from the out-of-time sample in the $t_{2max}$ control region. This
extrapolation leads to an estimate of $6.8 \pm 2.9$ events for both
in-time and accidental pile-up backgrounds. The uncertainty takes into
account the statistical error of the data in the control region as
well as the uncertainty of the extrapolation. This measurement 
is in good
agreement with previous considerations and, since it contains
the least number of assumptions, 
it is used for the branching
ratio calculation. As can be seen from Fig.~\ref{fig:rcplot}, the
$\ksppd$ background is not double-counted in this subtraction 
because it populates the
$R_C<7$ cm region.

Recently, the $\kspee$ decay has been observed for the first time.
Using the measured branching ratio~\cite{kspee} and taking into account the
cutoff on the invariant mass of the $e^+e^-$ pair $m_{ee}^2/m_K^2~>~0.2$ this
background has been estimated to contribute $ 0.6 \pm 0.3$ events.

\section{\bf Result}

\begin{figure}[ht]
\begin{center}
\mbox{\epsfig{file=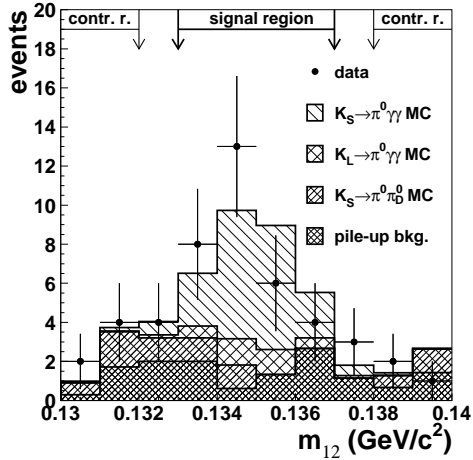,width=6.5cm}}
\end{center}
\caption{The invariant mass distribution of the photon pair assigned
to the $\pi^0$. The data in the control region are compatible with
background while the signal region contains a conspicuous excess
indicating a $\kspgg$ signal.}
\label{fig:m12plot}
\end{figure}

Of the selected 31 events, $13.7\pm 3.2$ are estimated to be
background (Table~\ref{syst}).
The probability to observe 31 or more events
in absence of a signal
with a background of $13.7 \pm 3.2$ events 
is $1.5 \times 10^{-3}$.
Subtraction of the background from the selected sample leads to a signal
of $17.3\pm 6.4$ events.
In the control region, defined as
$3<|m_{12}-m_{\pi^0}|<5$ MeV, the observed data agree well with
the background estimate (Fig.~\ref{fig:m12plot}). 

\begin{table}[htb]
\begin{center}
\begin{tabular}{|l|r|r|}
\hline
Events in & \multicolumn{1}{c|}{\parbox{1.3cm}{signal \\ region}} & 
            \multicolumn{1}{c|}{\parbox{1.3cm}{control \\ region}} \\ \hline        
$\klpgg$ background & 3.8  $\pm$ 0.2 & 0.1 $\pm$ 0.0 \\
$\ksppd$ background & 2.4  $\pm$ 1.2 & 4.2 $\pm$ 1.6 \\
Hadronic background & 0.1  $\pm$ 0.1 & 0.1 $\pm$ 0.1 \\
Pile-up background  & 6.8  $\pm$ 2.9 & 4.3 $\pm$ 2.0 \\ 
$\kspee$ background & 0.6  $\pm$ 0.3 &  $\ll$ 0.1      \\ \hline
Total background    & 13.7 $\pm$ 3.2 & 8.7 $\pm$ 2.6 \\ \hline
Data                & 31 & 9 \\ \hline
\end{tabular}
\end{center}
\caption{Summary of background composition in both signal
  and control regions, the latter being defined as
  $3<|m_{12}-m_{\pi^0}|<5$ MeV, compared to data.}
\label{syst}
\end{table}

The detector acceptances of the $\kspgg$ decay and of the normalisation 
decay channel $\kspp$
were calculated using a full Monte Carlo simulation of the
detector based on GEANT~\cite{geant}\footnote{In order to speed up the
  simulation, electromagnetic showers
  are pre-generated and stored in a shower library.}. 
The $\kspp$ acceptance was
determined to be 18.6\%. 
The $\kspgg$ generator used 
the decay matrix element calculated by~\cite{epd} using \chpt. The
acceptance of the $\kspgg$ for $z>0.2$ is $7.2 \pm 0.3 \%$, where 
the uncertainty is
extracted from the comparison with a simulation using a pure
phase-space decay generator. The $\kspgg$ acceptance is smaller than that of
$\kspp$ mainly due to cuts on $z^*_{\pi^0}$ and energy asymmetry
defined in (\ref{zstar}) and (\ref{asym}).

Normalising the signal to the collected $\kspp$ sample and taking into
account the acceptances results in a ratio of decay widths
\beq
\frac{\Gamma(\kspgg)_{z>0.2}}{\Gamma(\kspp)} =
(1.57 \pm 0.51_{stat} \pm 0.29_{syst}) \times 10^{-7}
\eeq
where the systematic uncertainty takes 
into account the background subtraction and
acceptance calculation uncertainties.
This result is stable against variations of most relevant cuts. In
particular, the result does not change significantly by releasing the
cut on the AKL anti-counter which increases the pile-up background by
a factor 5, and when requiring no associated hit in a scintillator hodoscope
placed upstream of the LKr calorimeter, which reduces the $\ksppd$
background by a factor 3.
Using $BR(\kspp)$ from \cite{pdg}, one obtains
\begin{eqnarray}
BR(\kspgg)_{z>0.2} &=& (4.9 \pm 1.6_{stat} \pm 0.9_{syst}) \times 10^{-8} \\
           &=& (4.9 \pm 1.8) \times 10^{-8}
\end{eqnarray}
which agrees with the predictions of \cite{epd}.

\begin{figure}[ht]
\begin{center}
\mbox{\epsfig{file=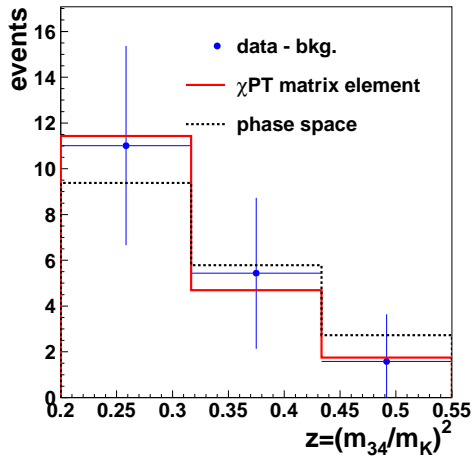,width=6.5cm}}
\end{center}
\caption{The $z$ distribution of data after background subtraction
  compared to a Monte Carlo $\kspgg$ simulation using phase space
  (dashed line) and \chpt\ matrix element~\cite{epd} 
(continuous line) in the decay generator.}
\label{fig:m34plot}
\end{figure}

In order to test the momentum dependence of the weak vertex predicted
by the \chpt\ the $z$ distribution of the sample after background
subtraction has been compared to $z$ distributions of simulated
$\kspgg$ events using two decay generators, one with the \chpt\ matrix element
and one with pure phase space. Fig.~\ref{fig:m34plot} shows that
both calculations agree within errors with the data and more
statistics would be needed to prove the chiral structure of the weak vertex.

\section*{ACKNOWLEDGEMENT}
It is a pleasure to thank the technical staff of the participating
laboratories, universities and affiliated computing centres for 
their efforts in the construction of the NA48 apparatus, in the 
operation of the
experiment, and in the processing of data. We would also like to thank
G.~D'Ambrosio for useful discussions.

\end{document}